\documentclass{aa}
\usepackage{tabularx}

\usepackage{graphicx}
\usepackage{txfonts}
\usepackage{hyperref}
\usepackage{ulem}

\graphicspath{{Images/}{../Images/}}
\usepackage{xcolor}

\begin{document}

    \title{The Possibility of Hydrogen-Water Demixing in Uranus, Neptune, K2-18\,b and TOI-270\,d
    }

    \author{S. Howard \inst{1}
          \and R. Helled\inst{1}
          \and A. Bergermann\inst{2,3}
          \and R. Redmer\inst{2}
          }

    \institute{Institut für Astrophysik, Universität Zürich, Winterthurerstr. 190, CH8057 Zurich, Switzerland,\\
              \email{saburo.howard@uzh.ch}
    \and 
    Institut für Physik, Universität Rostock, D-18051 Rostock, Germany
    \and
    SLAC National Accelerator Laboratory, 2575 Sand Hill Road, Menlo Park, California 94025, USA
              }
         
    \date{}
    
 
  \abstract
   {The internal structures of Uranus and Neptune remain unknown. In addition, sub-Neptunes are now thought to be the most common type of exoplanets. Improving our understanding of the physical processes that govern the interiors of such planets is therefore essential. Phase separation between planetary constituents may occur, in particular, hydrogen-water immiscibility in cold, water-rich intermediate-mass planets.
   }
   {We assess whether hydrogen-water demixing could occur in Uranus, Neptune,  K2-18\,b and TOI-270\,d, and investigate its effect on the planetary evolution and inferred internal structure.
   }
   {We couple planetary evolution models with recent \textit{ab initio} calculations of the hydrogen-water phase diagram, allowing for temperature shifts to account for uncertainties in miscibility gaps. 
   }
   {We find that demixing may occur and could lead to a complete depletion of water in the outermost regions of Uranus and Neptune. Temperature offsets of up to 1100~K lead to a depleted region comprising as much as 16\% of the planet's mass, and an increase in planetary radius by nearly 20\%. For K2-18\,b, our models suggest that hydrogen-water demixing is ongoing and may explain the absence of water features in its JWST spectrum. A temperature offset of 500~K is required to get a complete depletion of water in the atmosphere of K2-18\,b. TOI-270\,d may also have experienced hydrogen-water demixing. When applying a similar temperature offset on the phase diagram as for K2-18\,b, we find a partial depletion of water in the atmosphere of TOI-270\,d, consistent with JWST's detection of water.
   }
   {Hydrogen-water immiscibility may play a key role in shaping the structure and evolution of both Solar System giant planets like Uranus and Neptune, and cold/temperate exoplanets such as K2-18\,b and TOI-270\,d. Accounting for such internal processes is crucial to accurately interpret atmospheric observations from current (e.g., JWST) and upcoming (e.g., ARIEL) missions. 
   }

   \keywords{planets and satellites: gaseous planets --
                planets and satellites: interiors --
                planets and satellites: composition
               }

   \maketitle
%
\section{Introduction}

Revealing the internal structure of planets is crucial to understanding their formation pathways. Gravity data typically constrain static interior models of planets in the Solar System, whereas for exoplanets we use mass, radius, and, when available, atmospheric composition. Such models provide information about the present-day structure and composition but often assume \textit{a priori} different layers in the structure. Evolution models provide a complementary perspective as they capture how internal physical processes shape the structure over time and offer insights into the possible existence of distinct layers. Moreover, modeling the thermal evolution is fundamental as it links planetary formation and present-day internal structure. 

Key physical processes incorporated in evolution models include convective mixing, which can redistribute materials as planet evolves. This has been integrated into evolution models of Jupiter and Saturn \citep{vazan2018,muller2020dil,bodenheimer2025}, Uranus and Neptune \citep{vazan2020,arevalo2025_uranus}, and gas giant exoplanets \citep{knierim2024}. Another relevant process, the rainout of silicate triggered by their condensation, has been investigated in the context of sub-Neptune planets \citep{vazan2023,vazan2024}. Furthermore, phase separation between chemical species can drive internal differentiation and significantly impact planetary evolution.

For Jupiter and Saturn, it is now quite established that hydrogen-helium phase separation leads to helium rain, enriching the deep interiors with helium while depleting the outer layers. The pioneering work of \citet{smoluchowski1967,salpeter1973,stevenson1977b} allowed recent models \citep{mankovich2020,howard2024} to confirm that helium rain plays a key role in explaining the differences in the internal structures of Jupiter and Saturn.

The phase separation between hydrogen and water could affect the evolution of Uranus and Neptune. Interior models by \citet{bailey2021} suggested that Uranus may be in a more advanced stage of hydrogen-water demixing than Neptune, implying a current higher water content in Neptune's atmosphere and potentially explaining the disparity in heat flow between the planets. However, their analysis relied on the extrapolation of immiscibility curves beyond the range of available experimental data (up to 3~GPa) \citep{seward1981,bali2013}. Recently, \citet{cano2024}, followed \citet{bailey2021} and constructed interior models using updated experimental and theoretical data on the hydrogen-water phase diagram \citep{bergermann2021,vlasov2023,bergermann2024}. However, both \citet{bailey2021} and \citet{cano2024} focused exclusively on static models and did not consider their thermal evolution. 

In this paper, we incorporate hydrogen-water demixing into evolution simulations of Uranus and Neptune. 
Hydrogen-water demixing can occur when planets are sufficiently cold. Exoplanets may also be subject to this phenomenon, provided they lie within an appropriate temperature regime and contain enough water. We therefore also investigate the evolution of the exoplanets K2-18\,b and TOI-270\,d. 

Our paper is organized as follows. Section~\ref{sec:methods} explains our methods. Section~\ref{sec:solarsystem} presents our results for Uranus and Neptune, while Sec.~\ref{sec:exoplanets} presents our results for K2-18\,b and TOI-270\,d. Section~\ref{sec:discussion} discusses the limitations and future improvements of our work. Our conclusions are summarized in Sec.~\ref{sec:conclusion}.

\section{Methods}
\label{sec:methods}

\subsection{Phase diagram}

\begin{figure}[h]
   \centering
   \includegraphics[width=\hsize]{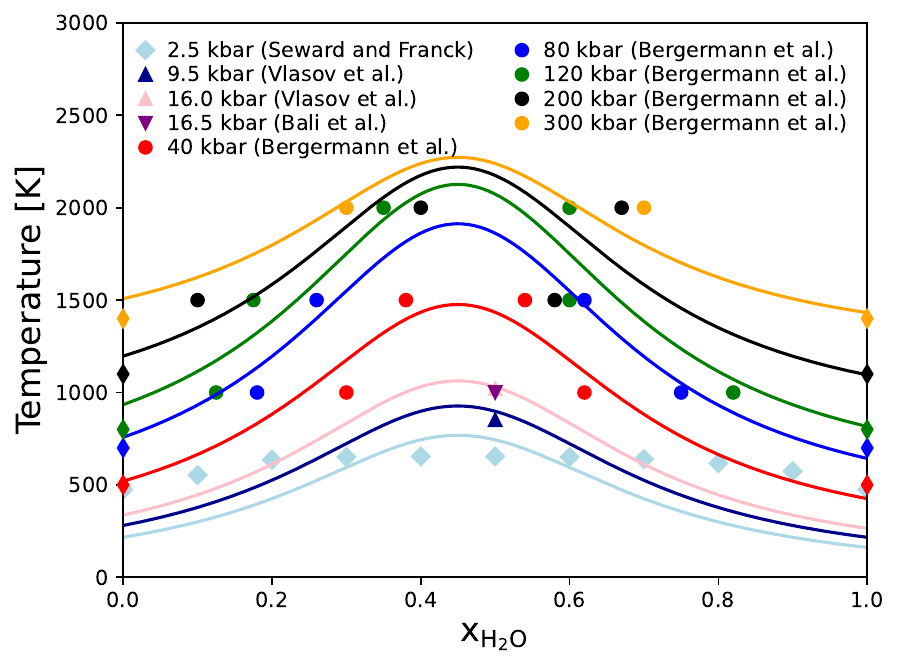}
      \caption{Miscibility gap of hydrogen and water for different temperatures (color-coded). DFT-MD simulations by \citep{bergermann2024} are shown as colored circles while earlier experimental campaigns of \citet{seward1981}, \cite{bali2013}, and \citet{vlasov2023} are presented by colored diamonds and traingles, respectively.}
         \label{figure:Armin_fit}
\end{figure}

The latest miscibility calculations using a combination of density functional theory and molecular dynamics (DFT-MD) of the mixtures $\rm H_2$ - $\rm H_2O$ could reach pressures up to 300~kbar, an order of magnitude higher than the maximal pressure obtained experimentally before \citep{bergermann2024}. Combining these ab initio simulations with the experimental data from \citet{seward1981,bali2013,vlasov2023} and the melting line of water, we fit a Lorentzian-like equation to interpolate the demixing temperatures for a wider range of compositions. The fit formula is discussed in Appendix~\ref{app:fit}. Next, we interpolated the curves over a larger range of pressures. This phase diagram will then be coupled with our planetary evolution models. Note that the $P-T-x$ conditions as shown in Fig.~\ref{figure:Armin_fit} are subject to uncertainties: The DFT-MD simulations might suffer from inaccuracies due to the choice of the exchange-correlation functional, while some of the experimental campaigns did not investigate the influence of the water concentration. Nevertheless, \citet{bali2013} and \citet{vlasov2023} used a similar experimental setup and found consistent results for most of the $P$-$T$ conditions.

In the following, we introduce $T_{\rm offset}$ which shifts the demixing diagram to higher temperatures while keeping the concentration and corresponding pressures constant. This shift is justified by uncertainties in the miscibility gap itself, uncertainties in the equation of state (EOS) (for further discussion see Appendix~\ref{app:waterEOS}), and missing knowledge of the exact planetary composition. Additionally, the shift allows us to investigate the influence of the miscibility gap for a wider range of scenarios.

\subsection{Model assumptions}
Our planetary evolution models were calculated with CEPAM \citep{guillot1995_cepam}, which solves the standard structure equations \citep[see, e.g.,][]{helledhoward2024}. These models build upon our previous work \citep{howard2024}, which focused on the evolution of Jupiter and Saturn, including hydrogen-helium phase separation. A key difference in the present study is the treatment of the energy equation: We adopt a formulation based on internal energy rather than entropy. This choice ensures the inclusion of a term that accounts for changes in composition \citep{strittmatter1970,morel1997}.

Our evolution models, whether for Uranus, Neptune, or K2-18\,b, initially have the following structure: a central core purely made of rocks, surrounded by an envelope consisting of a mixture of $\rm H_2$ - $\rm He$ - $\rm H_2O$. This envelope consists of a composition gradient in the water content, while the hydrogen-to-helium ratio is fixed to the protosolar value \citep{asplund2021}. This structure is roughly consistent with predictions of formation models \citep{valletta2022}. Basically, the parameters of our models are the mass of the core ($M_{\rm core}$), the water-mass fraction in the inner and outer regions of the envelope ($Z_{\rm H_2O,deep}$ and $Z_{\rm H_2O,atm}$, respectively), and the extent of the water gradient expressed as a normalized mass coordinate ($m_{\rm dilute}$) (as defined in \citet{howard2023_interior}). The slope of the water gradient was defined with $\delta m_{\rm dil}$ which was fixed to 0.075 in all our models. The temperature gradient in the envelope is assumed to be adiabatic. This assumption is discussed in Sec.~\ref{sec:discussion}. Given its higher mean density ($2.70~\rm g.cm^{-3}$), TOI-270\,d is likely to have a more rock-dominated interior. We hence adopt a simpler structure for this planet, which is discussed in Sec.~\ref{subsec:TOI-270d}.

We used the analytical EOS of \citet{hubbard1989} to compute the pressure-dependent density in the isothermal rocky core. For hydrogen and helium, we use the CMS19 + HG23 EOS \citep{chabrier2019,howardguillot2023}, and for water, we use QEOS \citep{more1988}. Note that more recent water EOSs exist. However, for simplicity, we use QEOS. Moreover, we calculated adiabats using QEOS and compared our findings with the results of \citep{cano2024} obtained using the AQUA EOS. Note that, as mentioned in \citet{aguichine2024}, there is a problem with the entropy in AQUA due to an error in the code of \citet{mazevet2019}. The comparison, presented in Appendix~\ref{app:waterEOS}, shows that the temperature difference due to the choice of water EOS ranges from about 500 to 1000~K between tens and 300~kbar. Such differences support the idea of shifting the temperature of the phase diagram. 

We use the atmospheric models of \citet{fortney2011} for Uranus and Neptune; and of \citet{guillot2010} for K2-18\,b and TOI-270\,d. The equilibrium temperatures are set to 58.1, 46.4, 254 and 354~K, respectively.

The demixing procedure used in this work follows the approach described in \citet{howard2024}, itself inspired by previous studies \citet{nettelmann2015,mankovich2016}, and has been adapted here for the demixing of $\rm H_2$ - $\rm H_2O$. This accounts for the presence of helium as our models feature ternary mixtures ($\rm H_2 - He - H_2O$) while the used phase diagram provides mass fractions corresponding to a binary mixture. We briefly summarize the procedure (further details can be found in \citet{howard2024}). As the planet cools, its $p$-$T$ profile may intersect the immiscibility region of the $\rm H_2$ and $\rm H_2O$ phase diagram. In the region of the planet where phase separation occurs, water droplets form and sink. Hence, such a region is depleted in water until it reaches saturation, that is, a new equilibrium composition where no further separation takes place. As the planet continues to cool, there might be a new equilibrium abundance, leading to further water depletion. In addition, the upper layers become progressively depleted, as convective mixing transports water downward to the region below the phase separation. Subsequently, the water is redistributed downwards, enriching the deeper layers. It is initially deposited in the layer directly beneath the region undergoing phase separation. 

Since our models feature an initial compositional gradient in water, and to ensure that the water distribution remains monotonic with depth, we apply an iterative procedure: if the underlying layer has a higher water abundance than the layer below it, we redistribute uniformly the removed water to the next layer down as well. This process is repeated until a monotonic water profile is achieved. Note that our models do not consider convective mixing and the settling after phase separation. 

\section{Application to Uranus and Neptune}
\label{sec:solarsystem}

\subsection{Uranus}
\label{subsec:uranus}

For Uranus, we present three evolution models to explore the effects of hydrogen–water demixing. Given the uncertainties in both the hydrogen-water phase diagram and the planet's temperature-pressure profile partly due to limitations in the water EOS, we allow for a temperature offset on the phase diagram. These three models correspond to different temperature offsets, i.e. $T_{\rm offset}=0$, 600 or 1100~K. The objective of the analysis is to assess the effect of hydrogen-water demixing, should it occur in the interior of the planet. Therefore, we did not conduct an extensive parameter study and did not aim to find the best Uranus model. Instead, we assess how materials are redistributed due to hydrogen-water demixing and investigate how the planetary radius and the temperature structure are affected.

\begin{figure*}
   \centering
   \includegraphics[width=0.87\hsize]{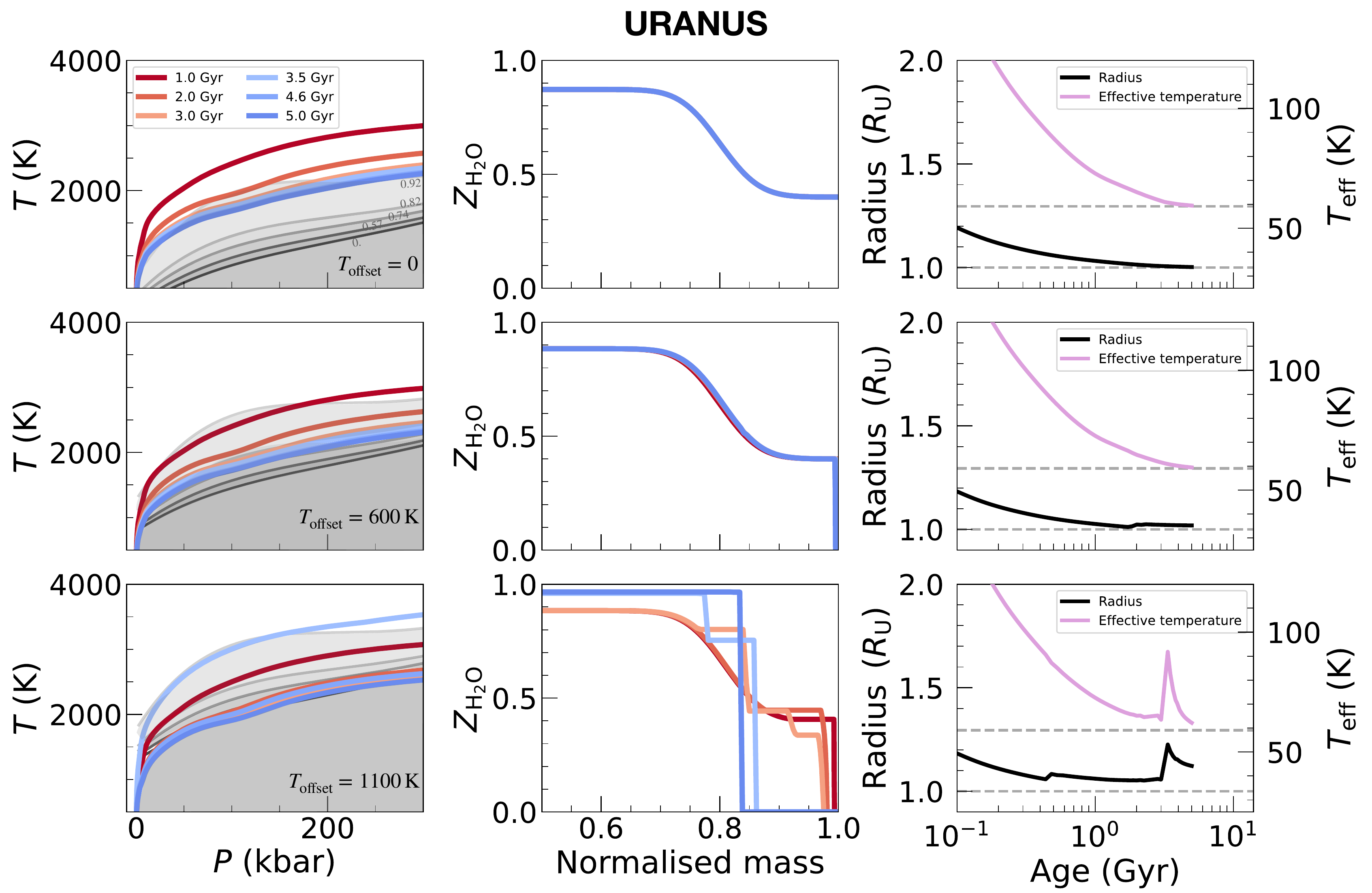}
      \caption{Evolution models of Uranus. Each row shows results for a different temperature offset applied to the hydrogen-water phase diagram ($T_{\rm offset}=0$, 600 or 1000~K.) \textit{Left panels}: temperature-pressure profiles at different ages. Phase curves are shown with grey to black lines for water mass fractions of 0.92 to 0. Grey-shaded areas mark demixing regions. \textit{Center panels}: water mass fraction as a function of normalized mass, colored by age. \textit{Right panels}: Radius and effective temperature as a function of age. The horizontal dashed lines show the measured values.
      }
         \label{figure:Uranus}
\end{figure*}

Figure~\ref{figure:Uranus} shows the results for Uranus. The top row corresponds to the case with $T_{\rm offset}=0$, i.e. with the original phase diagram (depicted with grey to black lines). We find that even after 5~Gyr of evolution, Uranus' adiabat is still too warm to intersect with demixing curves that correspond to water mass fractions lower than those present in Uranus. Therefore, no hydrogen-water demixing occurs and the initial water profile remains unchanged throughout the entire evolution. This model fits the observed mean radius and effective temperature because the parameters were optimized for that, using a Monte-Carlo method \citep[see][]{muller2023}. The parameters of the model are: $M_{\rm core}=4~M_{\oplus}$, $m_{\rm dilute}=0.8$, $Z_{\rm H_2O,deep}=0.88$ and $Z_{\rm H_2O,atm}=0.4$. 

The middle row shows the results for $T_{\rm offset}=600~$K. In this case, Uranus' adiabat cools sufficiently to experience hydrogen-water demixing. After 2~Gyr, demixing begins but only in the outermost region of the planet, at pressures below 2.4~kbar. This represents only a few percent of the planet's mass. Thus, the abundance of water $Z_{\rm H_2O,atm}$ abruptly drops to zero. 
The relatively small amount of water removed in the outer regions and redistributed deeper affects the planetary radius. At  $\sim$2~Gyr, the radius slightly increases (by 3\%) driven by the release of energy due to demixing. This radius inflation was already predicted in \citet{bailey2021}, as well as in \citet{vazan2024} who studied the settlement of silicates. We note that adjusting the initial parameters to yield a denser structure would allow the model to match the observed radius at the end of the evolution.

Lastly, we present a case where $T_{\rm offset}=1100~$K, in which Uranus has "fully crossed" the hydrogen-water phase diagram. Increasing the temperature offset causes the demixing process to occur earlier. The onset of demixing found at $\sim$2~Gyr for $T_{\rm offset}=600~$K now occurs at 500~Myr. A similar decrease of $Z_{\rm H_2O,atm}$ to 0 and an increase of the planetary radius is found. As the planet continues to evolve, a progressively larger outer portion of the planet is being depleted in water. At about 3~Gyr, Uranus' adiabat intersects with the hydrogen-water phase diagram such that the outer 15\% of the planet by mass undergo demixing. Because only a few percent were affected previously, this now results in a substantial radius increase of 20\%. The effective temperature is also significantly affected and increases by 43\%. The bottom left panel shows that Uranus' adiabat heats up to temperatures even higher than at earlier times. Eventually, Uranus becomes fully demixed and its water structure becomes differentiated, leaving an outer region (corresponding to the outer 16\% in mass) completely depleted in water, and an inner region enriched in water, where $Z_{\rm H_2O,deep}$ is close to 1. 

Finally, as shown in Appendix~\ref{app:J2J4}, the redistribution of material due to hydrogen-water demixing significantly affects the calculated gravitational moments ($J_2$ can typically be affected by the order of $\sim 1000~$ppm). Furthermore, it was shown that static models with $Z_{\rm H_2O,atm} \sim 0$ can fit the observed gravity field \citep{nettelmann2013,morf2024}. Models with such abundance in atmospheric water are also part of the range of solutions found by \citet{bailey2021}. However, \citet{cano2024} find a lower limit of 0.05, although the region where phase separation occurs (between 40 and 110~kbar) is in line with what we obtain.

\begin{figure*}[!t]
   \centering
   \includegraphics[width=0.87\hsize]{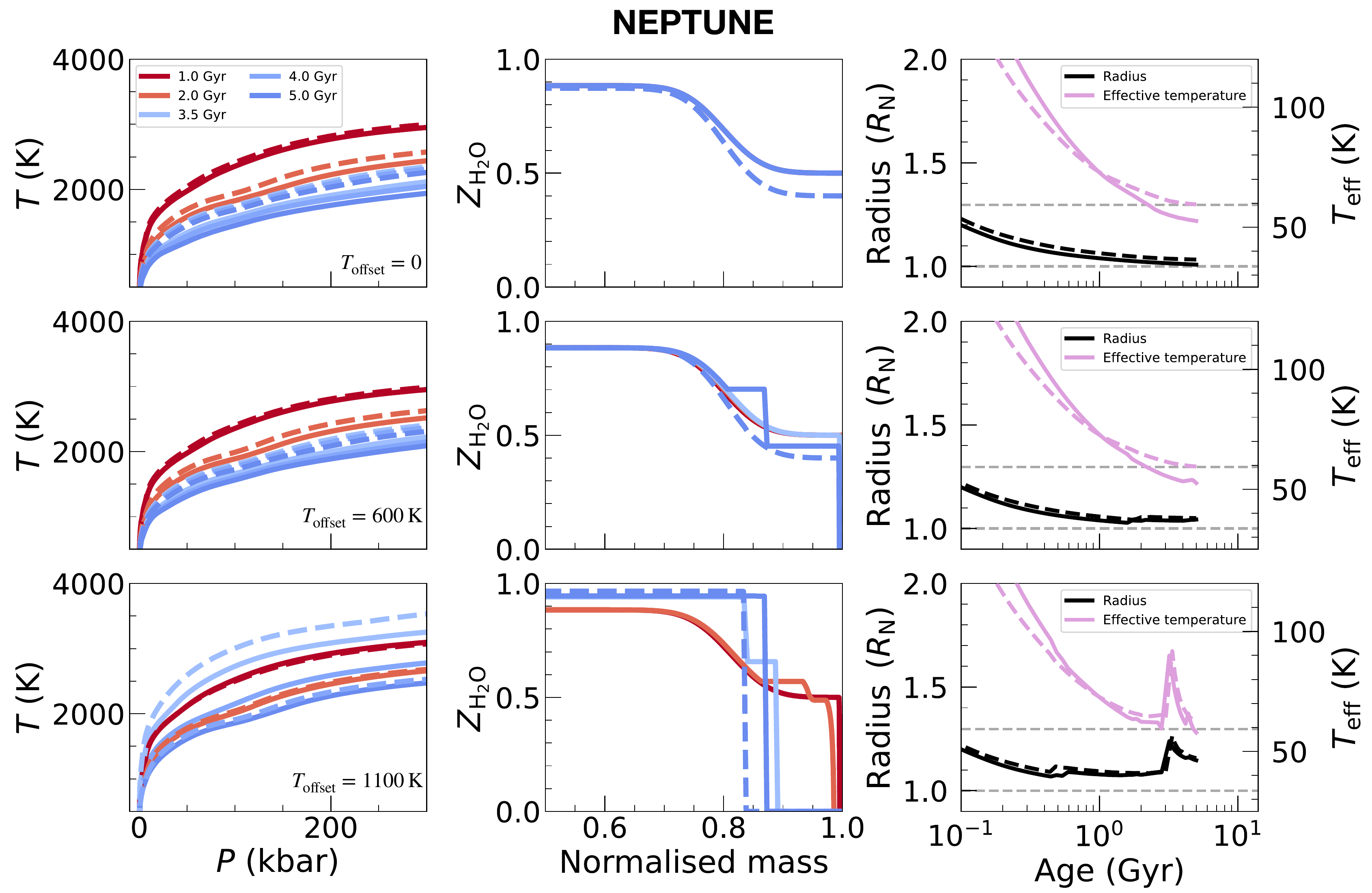}
      \caption{Evolution models of Neptune. Description is same as for Fig.~\ref{figure:Uranus}. For comparison, Uranus models are shown with dashed lines.
      }
         \label{figure:Neptune}
\end{figure*}

\subsection{Comparison with Neptune}
\label{subsec:neptune}

For Neptune, we also present three evolution models, using the same temperature offsets as for Uranus. The only modification is an adjustment of the initial water mass fraction in the outer envelope, which we set to $Z_{\rm H_2O,atm}=0.5$ instead of 0.4. This change allows the model with $T_{\rm offset}=0$ to reproduce Neptune's observed radius, which is slightly smaller than that of Uranus. Aside from this adjustment, the parameterization remains similar to that used for Uranus.

In the case with $T_{\rm offset}=0$, no hydrogen-water demixing occurs, as with Uranus. The model matches Neptune's observed radius. However, the resulting effective temperature is 11\% lower than the observed value. A dedicated parameter study would be required to identify models that accurately reproduce both the radius and effective temperature. We find that Neptune's interior is cooler than that of Uranus at the end of the evolution, which is expected due to the lower outer boundary temperature and the assumption of a similar compositional setup with an adiabatic temperature gradient. Under these comparable conditions, we anticipate that Neptune may be at a more advanced stage of hydrogen-water demixing.

When using $T_{\rm offset}=600~$K, demixing occurs and is more pronounced in Neptune than in Uranus, owing to its colder temperature-pressure profile. As with Uranus, the outermost layers -- representing only a few percent of the planet's mass -- become fully depleted in water. Additionally, the primordial water gradient begins to erode due to the demixing process. Indeed, the outer 13\% of the planet by mass (initially with $Z_{\rm H_2O,atm}=0.5$) experiences water depletion, slightly decreasing to $Z_{\rm H_2O,atm}=0.44$. Nevertheless, the resulting increase in planetary radius remains comparable to that of Uranus.

Using $T_{\rm offset}=1100~$K leads to a fully demixed Neptune, as also inferred for Uranus. The associated increases in planetary radius and effective temperature are of similar magnitude between the two planets. The main distinction lies in the extent of the completely depleted region in water: the outer layer where $Z_{\rm H_2O,atm}=0$ comprises 13\% of the Neptune mass, compared to 16\% for Uranus.

Here, we compare evolution models of Uranus and Neptune using similarly shifted hydrogen-water phase diagrams. We found that Neptune's interior is colder than that of Uranus, which naturally arises from its lower equilibrium temperature and the assumption of broadly similar internal structures. Under these conditions, Neptune is expected to be at a more advanced stage of demixing for a given immiscibility diagram. This may be consistent with the results by \citet{cano2024} finding $Z_{\rm H_2O,atm}$ ranging from 0.05 and 0.21 for Uranus and between 0.05 and 0.16 for Neptune. However, \citet{bailey2021} suggested the opposite trend based on the significant difference in intrinsic heat flux between the two planets. Indeed, the ratio of total emitted to absorbed solar power is low for Uranus, compared to Neptune (and Jupiter and Saturn) \citep{irwin2025,wang2025}. Our simple comparison does not resolve the long-standing debate about this heat flux disparity. It is possible that Uranus and Neptune differ since or due to the formation, with variations in their bulk composition and structure (such as core mass or compositional gradients) and distinct thermal structures, which could lead to different effects of convective mixing. These differences may significantly influence their internal temperature profiles and, consequently, the degree and characteristics of hydrogen-water demixing within each planet. Nevertheless, Neptune's outer envelope could still be fully depleted in water, similar to Uranus. Such a scenario would require an enrichment in other volatiles to be consistent with static models that suggest a higher heavy-element content in Neptune's envelope compared to that of Uranus, and the observed heat flux disparity to be primarily attributed to differences in convective mixing rather than demixing alone.

\section{Application to exoplanets}
\label{sec:exoplanets}

\subsection{K2-18\,b}
\label{subsec:K2-18b}

K2-18\,b has recently gained interest due to its potential habitability, sparking ongoing discussions \citep{madhu2020,madhu2023,shorttle2024,wogan2024,schmidt2025,madhu2025,taylor2025,welbanks2025,pica2025,luque2025}. Below, we show that this planet can also undergo hydrogen-water demixing. 
With a mass of $8.63 \pm 1.35~M_{\oplus}$ \citep{cloutier2019} and a radius of $2.61 \pm 0.09~R_{\oplus}$ \citep{benneke2019}, K2-18\,b has an equilibrium temperature of 254~K. Moreover, its mean density ($2.67~\rm g.cm^{-3}$) which is more than twice that of Uranus, suggests that the interior is more enriched with heavy elements. A higher heavy-element content leads to colder adiabats, as illustrated by the comparison between adiabats with $Z_{\rm H_2O}=0.4$ and 0.8 in Fig.~\ref{figure:waterEOS}.

Observations from the Hubble Space Telescope (HST) suggested the presence of $\rm H_2O$ in the atmosphere of K2-18\,b \citep{madhu2020}. However, the James Webb Space Telescope (JWST) did not detect water \citep{madhu2023,schmidt2025}, which was attributed to condensation and cold trapping. Here, we show that the absence of water could be due to hydrogen-water demixing.

Various scenarios have been proposed for the internal structure of K2-18\,b. \citet{madhu2020} initially suggested three of them: a rocky world, a water world, and a mini-Neptune. Photochemical and climate considerations have been put forward to try to distinguish the different scenarios. Whereas \citet{madhu2023} favour the water (or hycean) world interpretation, other studies \citep{shorttle2024,wogan2024,schmidt2025} suggest that a simpler mini-Neptune structure is sufficient to explain the observations. Here, we only consider radius, mass, age, and equilibrium temperature as constraints of our evolution models. 
The question of habitability is not the focus of our work. 
Our primary objective is to investigate the potential occurrence of hydrogen-water demixing in a planet like K2-18\,b. However, we include a water phase diagram to assess whether liquid water could be present in Appendix~\ref{app:liquid}.

\begin{figure}[h]
   \centering
   \includegraphics[width=0.7\hsize]{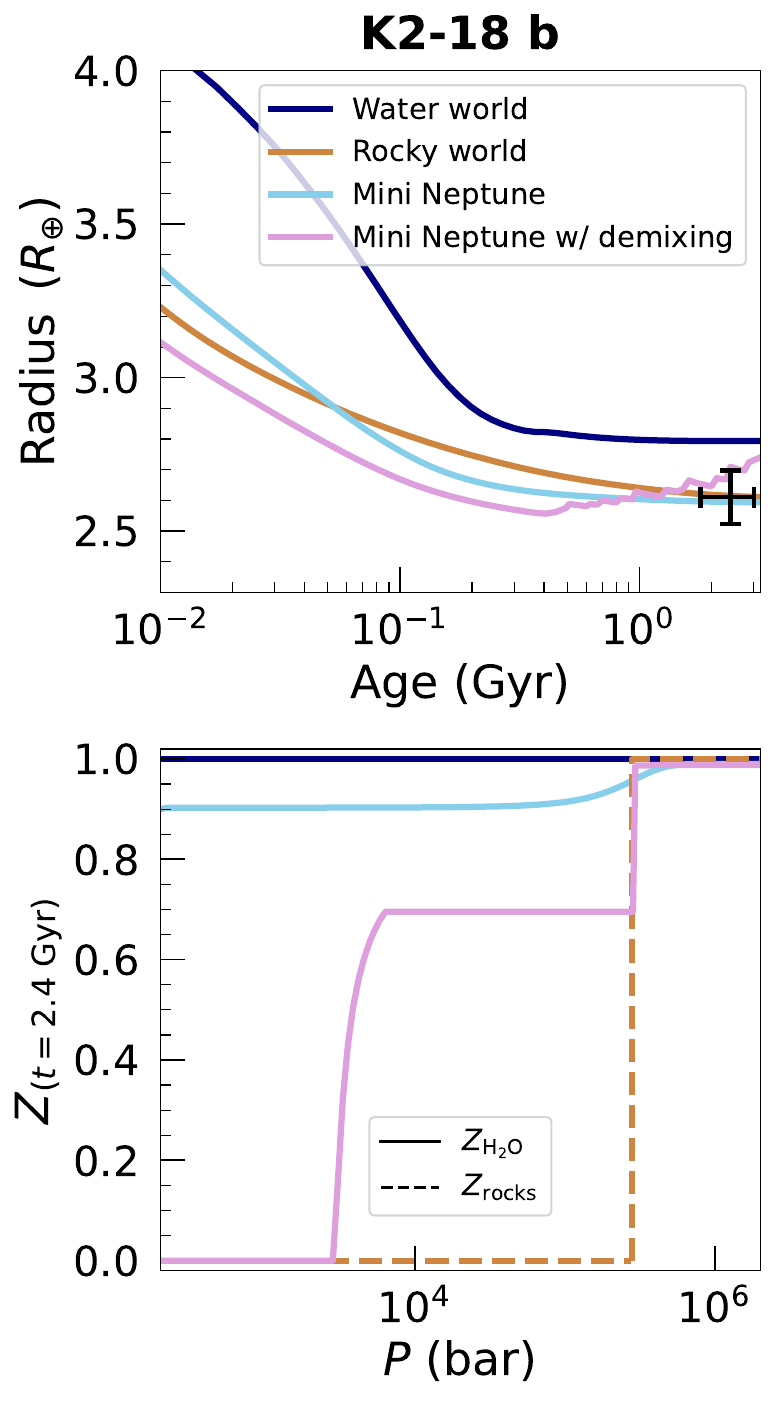}
      \caption{Evolution models of K2-18\,b. \textit{Top panel:} planetary radius as a function of age. The black errorbar corresponds to the measured radius and age. \textit{Bottom panel:} heavy-element mass fraction at 2.4~Gyr (the planet's age) as a function of pressure.
      We note that condensation was not considered in our models, which explains why some models have a high heavy-element mass fraction at low pressures.
      }
         \label{figure:K2-18b}
\end{figure}

We first consider the rocky world scenario. We model the interior with a rocky core that represents 94\% of the planetary mass, overlaid by a H-He envelope that represents 6\% of the mass. As shown in Fig.~\ref{figure:K2-18b}, we find that this model fits the observed radius, and it is perfectly in line with the rocky world model from \citet{madhu2020} as well as the "ice-poor" case from \citet{schmidt2025}.

We then consider the water-world scenario. Our model is only made of water (where $Z_{\rm H_2O} \sim 1$), with a minimal H-He atmosphere that represents $10^{-6}$ of the planetary mass, as in \citet{madhu2020}. However, such a model yields a radius that is higher than the observed value. {Note that we do not exclude the water-world scenario as the inferred radius depends on the considered material and its given EOS \citep[see e.g.,][]{valencia1013,howard2025}. A denser water EOS (such as AQUA for instance) could help fit the observed radius.

Finally, we consider the mini-Neptune scenario. We calculated two models: with and without hydrogen-water demixing. Without demixing, we model the planet with a core that represents 37\% of the planetary mass. We chose $Z_{\rm H_2O,atm}=0.9$, $Z_{\rm H_2O,deep}=0.99$ and $m_{\rm dilute}=0.9$. Such parameters allow to fit the observed radius. In this model, H-He represents less than 2\% of the mass and water is about 60\%. Such a model is included in the range of solutions proposed by \citet{madhu2020} and \citet{schmidt2025}. Next, we consider hydrogen-water demixing. We chose $T_{\rm offset}=500~$K, which is the minimum temperature offset required to fully deplete the planet's atmosphere in water. We present one model fitting the observed radius. The core represents 60\% of the planetary mass and the assumed parameters are $Z_{\rm H_2O,atm}=0.8$, $Z_{\rm H_2O,deep}=0.95$ and $m_{\rm dilute}=0.9$. Since demixing is expected to increase the planet's radius, this set of parameters allowed to make the planet denser compared to the case without demixing and hence helped to fit the observed radius. In this particular model, we found demixing starting after 400~Myr. At the planet's current age (2.4~Gyr), the outer region of the planet ($P<2.8~$kbar) is fully depleted in water. As differentiation is still ongoing, we find a region where $Z_{\rm H_2O}=0.7$, between 6 and 281~kbar. At higher pressures, the water mass fraction reaches 0.99. This mini-Neptune scenario including hydrogen-water demixing demonstrates that such a phenomenon can lead to significant depletion of water in the outer regions, potentially explaining the non-detection of $\rm H_2O$ in JWST observations of K2-18\,b's atmosphere. We emphasize that this is only one model including demixing. We have not conducted a full exploration of the parameter space; our aim is simply to test the plausibility of hydrogen-water demixing as a mechanism that could reconcile internal structure modeling with atmospheric observations. Alternative solutions could certainly exist, and we hope to investigate them in future research. 

\subsection{Comparison with TOI-270\,d}
\label{subsec:TOI-270d}

With a mass of $4.78 \pm 0.43~M_{\oplus}$ and a radius of $2.133 \pm 0.058~R_{\oplus}$ \citep{vaneylen2021}, TOI-270\,d is a temperate sub-Neptune ($T_{\rm eq}=354~$K) that also gained interest. Recently, \citet{benneke2024} suggested that, due to its slightly higher equilibrium temperature and lower mass compared to K2-18\,b, TOI-270\,d is less likely to be a hycean world. The inferred interior structure was found to consist of a rock/iron core that represents 90\% of the planet's mass, overlaid by an envelope with a heavy-element mass fraction of $Z_{\rm atm}=0.58^{+0.08}_{-0.12}$ (as indicated by JWST).

First, we do not consider hydrogen-water demixing and calculate a similar rock-dominated model, assuming a core that accounts for 90\% the planet's mass and an envelope with $Z_{\rm atm}=0.58^{+0.08}_{-0.12}$. Given the uncertainty in the planet’s age, we adopt an age range of 1 to 10~Gyr, as done in \citet{benneke2024}. We find that this model can reproduce the observed radius (see Fig.~\ref{figure:toi-270d}). Second, we include the possibility of demixing, using the same temperature offset of $T_{\rm offset}=500~$K as for K2-18\,b. In this case, demixing occurs but does not fully deplete the planet's atmosphere of water. The planetary radius increases by 1.5\% due to demixing but remains consistent with observational constraints. $Z_{\rm atm}$ is reduced from 0.58 to 0.39 and is not in line with the observed value. However, the atmospheric heavy-element mass fraction measured by JWST includes other ices (e.g., $\rm CH_4$, $\rm CO_2$) whereas our models consider only $\rm H_2O$. Future models that distinguish different types of ices would be valuable (see Sec.~\ref{sec:discussion}). However, this model shows that hydrogen-water demixing could occur and affect the evolution of TOI-270\,d.

\begin{figure}[h]
   \centering
   \includegraphics[width=0.75\hsize]{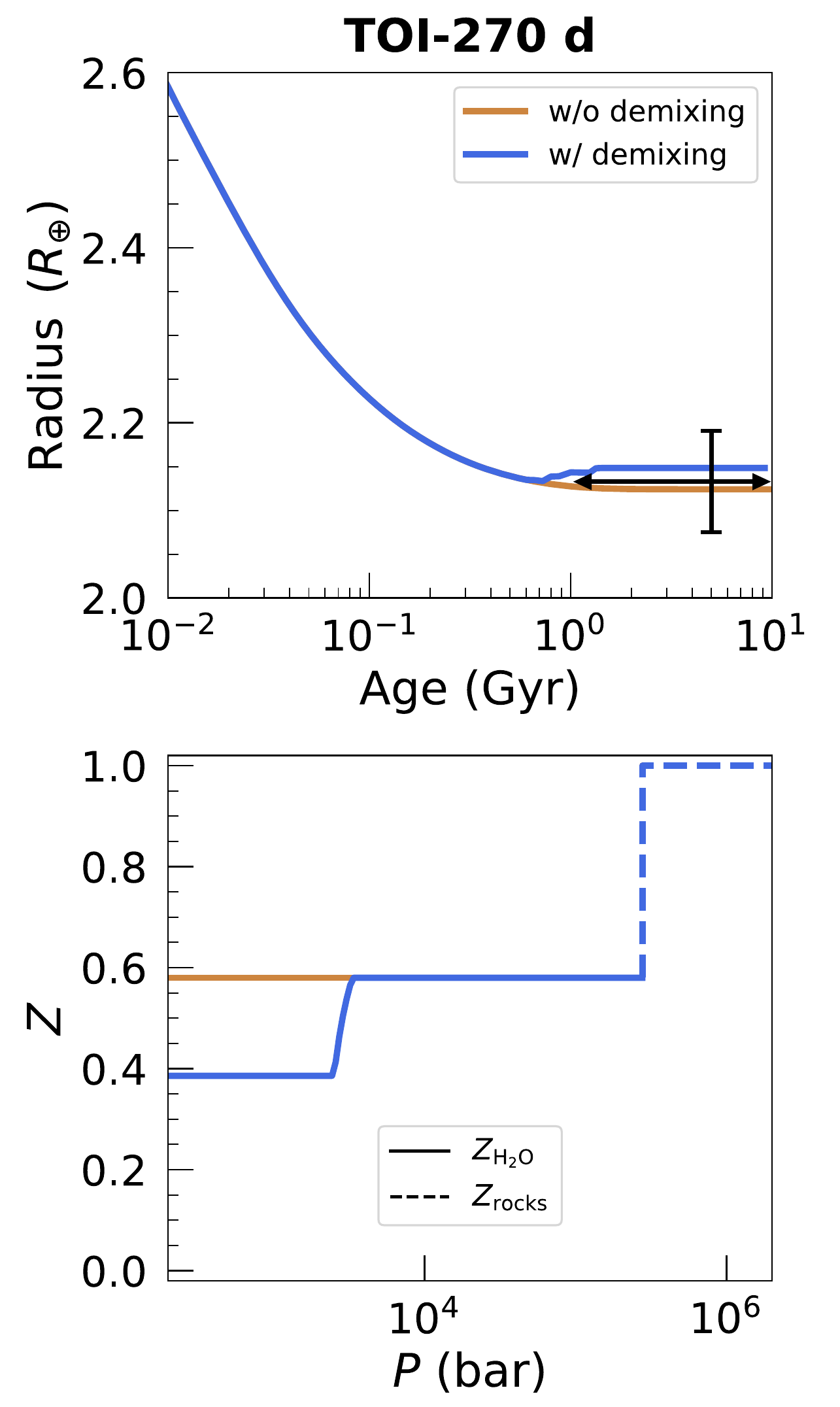}
      \caption{Evolution models of TOI-270\,d. \textit{Top panel:} planetary radius as a function of age. The black errorbar corresponds to the measured radius and age.  Since the age is uncertain, we adopted an age range of 1 to 10~Gyr. \textit{Bottom panel:} heavy-element mass fraction at the end of the evolution as a function of pressure.
      }
         \label{figure:toi-270d}
\end{figure}

Figure~\ref{figure:k2vstoi} compares the $P$-$T$ profiles of the envelopes of TOI-270\,d (between 1 and 10~Gyr) and K2-18\,b (between 500~Myr and 2.4~Gyr). We note that the temperature profile of TOI-270\,d does not evolve after 1~Gyr as it stopped cooling. This can be attributed to the presence of a large core, which quickens the cooling of the planetary envelope. This also explains why the radius evolution flattens (Fig.~\ref{figure:toi-270d}). The inset of Fig.~\ref{figure:k2vstoi} shows the sensitivity of our results to the low-pressure part (a few kbars) of the phase diagram (see Sec.~\ref{sec:discussion} for further details). The $T$--$P$ profiles of K2-18\,b are sufficiently cold to intersect the demixing curve that corresponds to a water mass fraction of 0. However, this is not the case for TOI-270\,d, and as a result, its outer envelope is not completely depleted in water.  
It is clear that other models may be possible for TOI-270\,d and we hope to address this in future research. 

\begin{figure}[h]
   \centering
   \includegraphics[width=0.95\hsize]{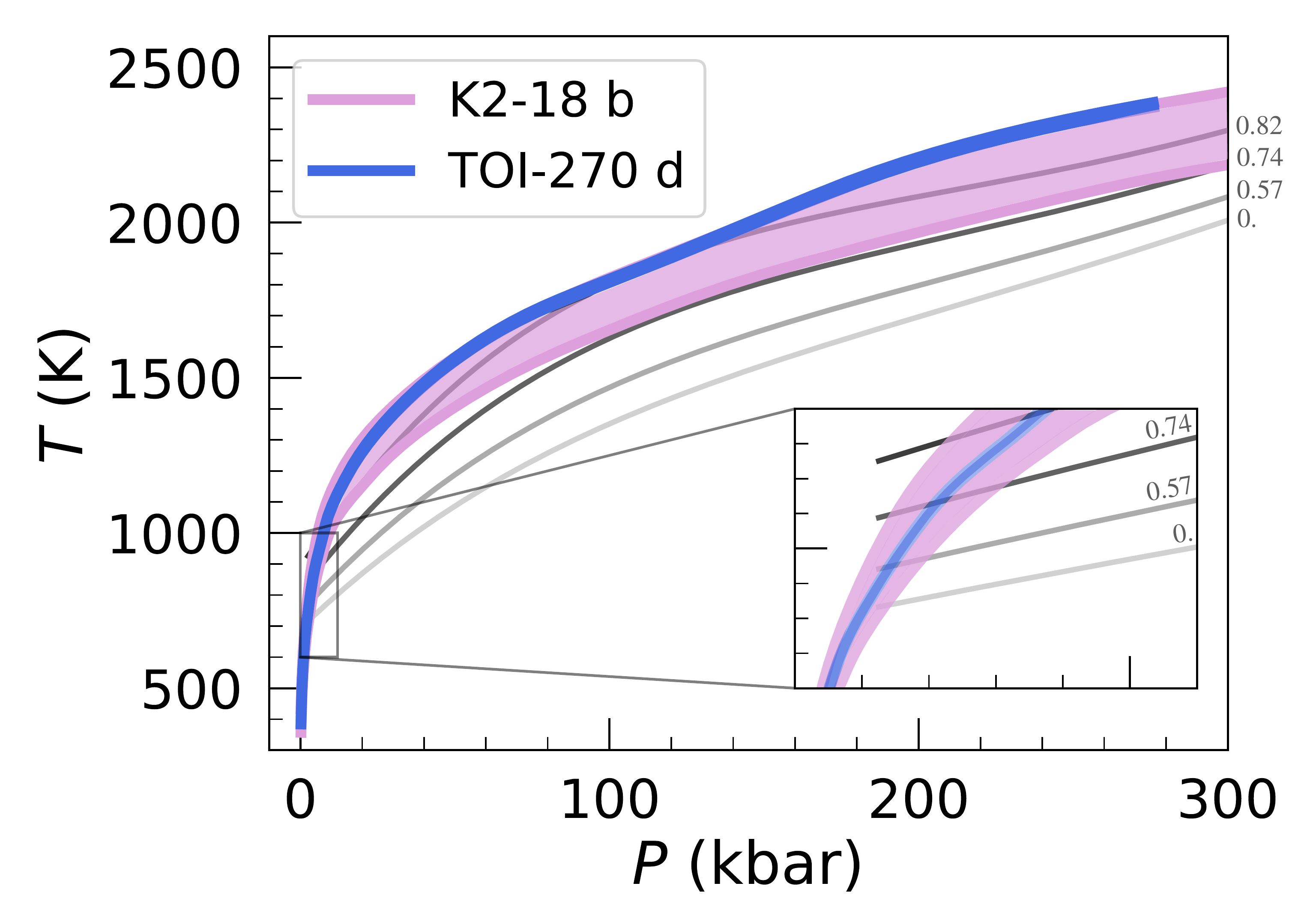}
      \caption{Temperature-pressure profiles of the envelopes of K2-18\,b and TOI-270\,d. A range of profiles between 500~Myr and 2.4~Gyr is shown for K2-18\,b while a range between 1 and 10~Gyr is shown for TOI-270\,d. Phase curves for different water mass fractions are shown with grey to black lines.
      }
         \label{figure:k2vstoi}
\end{figure}

\section{Discussion}
\label{sec:discussion}
Our results are based on several assumptions, some of which warrant further discussion. In this section, we discuss the limitations of our approach, highlight important aspects that merit further investigation, and propose potential improvements of the models presented here. \\

\textit{Shifting the phase diagram.} In our models, we considered a temperature shift in the  hydrogen-water phase diagram by applying an offset. An offset is possible given the uncertainties in the phase diagram.    
An additional motivation for applying such a shift is that the water EOS used here predicts lower densities compared to more recent models. Nevertheless, one may question whether applying a constant offset to the entire phase diagram is a valid approach. The low-pressure regime (at a few kbar) of the phase diagram -- and similarly of the EOS -- may be better known than the high-pressure regime. Since demixing begins in the outer layers of the planet, our results could be biased by applying the same shift to the whole phase diagram. However, in our Uranus model with $T_{\rm offset}=600~$K (Fig.~\ref{figure:Uranus}), demixing occurs at pressures below 2.4~kbar where temperature differences of 100-400~K still arise due to uncertainties in the water EOS (see Appendix~\ref{app:waterEOS}). Therefore, applying a temperature offset in the low-pressure regime is not unreasonable. Once the planet has fully crossed the phase diagram, the low-pressure region becomes less critical, as the adiabat intersects the demixing curves at higher pressures as well. This still results in depletion of the outer layers, as assumed in our demixing procedure. However, if demixing is still ongoing, then the low-pressure part of the phase diagram remains important, as do the planetary $T$-$P$ profiles. Further calculations of the hydrogen-water phase diagram would therefore be valuable. Improving our knowledge of planetary $T$--$P$ profiles would benefit from better constraints on composition and EOSs. In addition, other processes such as condensation, discussed later, may influence the thermal structure and should be investigated more in depth.

Concerning the phase diagram, the calculations from \citet{bergermann2024} reach up to 300~kbar. One may wonder what should happen at higher pressures. We considered this value as the upper limit of the phase diagram. Similarly to hydrogen and helium mixtures, we can expect hydrogen and water to be immiscible once one of the constituents becomes solid. This immiscibility could occur as soon as the phase transition from liquid to superionic water occurs \citep{gupta2025}.

\textit{Initial composition.} Our evolutionary models account for primordial composition gradients, representing a step beyond previous approaches. Formation models of Uranus and Neptune \citep{valletta2020,valletta2022} as well as models of mini-Neptunes \citep{ormel2021} suggest that such gradients are expected. Some interior models of Uranus that fit the measured gravity field also include such gradients \citep{benno2022,benno2024,morf2024,lin2025}. In this study, the initial composition profiles were chosen arbitrarily, primarily to be consistent with the observed radius. We did not conduct a comprehensive parameter study, and, undoubtedly, alternative solutions are possible. Future work should explore the extent and slope of the water gradient, as well as the initial deep and atmospheric water abundances. In particular, since we found that the abrupt drop of $Z_{\rm H_2O,atm}$ to 0 affects the planetary evolution, higher initial atmospheric water abundances could have an even greater effect. In our models, water was the only ice component considered. Including other components such as methane and ammonia could influence the miscibility gap of hydrogen and water. Additionally, we did not mix rocks and water, but assumed that the rocks are  concentrated in a central core. In reality, rocks and ices could mix \citep{vazan2022}. and this should be considered in future models. 
Moreover, phase separations other than between hydrogen and water should also be considered (e.g., H-He \citep{lorenzen2009,lorenzen2011,Morales2013b,schottler2018,schottler_jpp,brygoo2021,Bergermann2021a}, C-N-O \citep{militzer2024} but also C-H \citep{Cheng2023} or C-H-O \cite{He2022}). Future studies exploring the parameter space would be valuable to identify models that can reproduce the observational constraints, particularly after accounting for the radius inflation caused by demixing.

\textit{Temperature gradient.} In our models, we assumed an adiabatic temperature gradient throughout the envelope. However, the presence of composition gradients implies that the temperature gradient may deviate from adiabatic, though the degree of super-adiabaticity remains uncertain. If the deviation is small, the adiabatic assumption might still be a reasonable approximation. To assess the potential impact, we ran a Uranus model (where $T_{\rm offset}=600~$K, similar to the one presented in Sec.~\ref{subsec:uranus}) but assuming a super-adiabatic temperature gradient at the location of the primordial water gradient. We employ the parameterization used in \citet{nettelmann2015,mankovich2020,howard2024}:
\begin{equation}
    \nabla_{T}=\nabla_{\rm ad}+R_{\rho}B,
\end{equation}
where $B = \frac{\chi_{\rho}}{\chi_{\rm T}}\left(\frac{d\,\textrm{ln}\,\rho}{d\, \textrm{ln}\,Z} \right)_{P,T} \nabla_{Z}$, $\chi_{\rho} = \left(\frac{d\,\textrm{ln}\,P}{d\, \textrm{ln}\,\rho} \right)_{T,Z}$, $\chi_{\rm T} = \left(\frac{d\,\textrm{ln}\,P}{d\, \textrm{ln}\,T} \right)_{\rho,Z}$ and $\nabla_{Z}=\frac{d\,\textrm{ln}\,Z}{d\,\textrm{ln}\,P}$. We used $R_{\rho}=0.01$ yielding a central temperature hotter by 1500~K. The departure of the $T$--$P$ profiles from adiabatic ones are shown in Appendix~\ref{app:nonadiabatic}. We do not find a significant impact of the super-adiabatic gradient on the occurrence of demixing, nor on its effects on the planetary radius or effective temperature evolution. However, this result depends on the assumed value of $R_{\rho}$ and on the characteristics of the water gradient. In our model, the water gradient is located at pressures higher than those where phase separation occurs, which may limit its influence on the demixing process. In addition, previous evolution models of Uranus and Neptune have suggested that neither planet is fully adiabatic \citep{nettelmann2016,scheibe2019,scheibe2021}. Future studies should explore the parameter space of the temperature gradient -- not only at the location of the primordial water gradient but also in the region undergoing phase separation -- in order to better constrain its effect on the planetary evolution.

\textit{Condensation.} Our models do not consider condensation of water or other volatile species. Nevertheless, condensation remains a plausible explanation for the absence of detectable water in observed atmospheres. Studies of the atmospheric dynamics and cloud formation \citep[e.g.,][]{charnay2021} are essential to assess the role of condensation in shaping the thermal and compositional structure. Condensation could also inhibit convection, as discussed in \citet{guillot1995_condensation,leconte2017,markham2021}, and affect the temperature gradient in the planetary interior. This could, in turn, change the effect of hydrogen-water demixing on the evolution. Moreover, if condensation inhibits convection, it could hinder mixing and lead to a decoupling between the atmosphere and the outer envelope. In that case, the atmosphere would no longer be able to supply water to the region where phase separation occurs.

\section{Conclusion}
\label{sec:conclusion}

In this study, we coupled a planetary evolution model with recent \textit{ab initio} calculations of the hydrogen-water phase diagram \citep{bergermann2024}. We allowed for temperature shifts to account for uncertainties in miscibility gaps, and modelled the thermal evolution of Uranus, Neptune, K2-18\,b and TOI-270\,d including hydrogen-water demixing. 
We found that demixing could occur in their interiors and significantly affect their evolution and internal structure. 
Fig.~\ref{figure:sketchs} shows the inferred interiors of the planets if hydrogen-water demixing occurs. 
Our main results can be summarized as follows:

\begin{itemize}
    \item Hydrogen-water demixing may occur and could fully deplete the outermost region of Uranus and Neptune in water. Applying temperature offsets on the phase diagram of 600 or 1100~K, we find that this outermost region would represent either a few percent of the planet's mass or up to 16\%. It leads to an increase in the planetary radius by 3\% and 20\%, respectively.
    \item Hydrogen–water demixing could occur in K2-18\,b. This process might explain why the JWST did not detect water in its atmosphere. A mini-Neptune scenario considering this phenomenon can fit the observed radius, mass, age, and equilibrium temperature. A temperature offset of 500~K is required to completely deplete the water in the atmosphere of K2-18\,b.
    \item TOI-270\,d could also have experienced hydrogen-water demixing. When applying a similar temperature offset on the phase diagram as for K2-18\,b, we find a depletion of water in the atmosphere of TOI-270\,d but not a complete one. This is in line with the detection of water by JWST.
\end{itemize}

Our work highlights the importance of thermodynamic processes in shaping planetary interiors throughout their evolution. It is relevant for both solar system planets and exoplanets. If hydrogen-water demixing occurs in Uranus and Neptune, it should be accounted for in static models constrained by gravity data, as it can erode primordial water gradients and affect the calculated $J_2$ by $\sim 1000~$ppm. Future work should also incorporate constraints from dynamo models \citep{stanley2004,stanley2006}, as done in \citet{militzer2024}. Measurements of the atmospheric compositions of Uranus and Neptune would help constrain evolution models and possibly reconcile the observed heat-flux disparity between the two planets.

In the context of temperate and cold exoplanets, miscibility gaps -- such as between hydrogen and water -- must be considered. These processes can link the atmosphere to the deep interior, as material may be transported inward. With the remarkable quality of atmospheric data now available from instruments like JWST and the upcoming ARIEL mission, it is essential to better understand such interior processes in order to accurately interpret atmospheric spectra. This requires advances in planetary interiors and evolution models, as well as synergies with the high-pressure physics community, to refine phase diagrams and equations of state.

\begin{figure}[h]
   \centering
   \includegraphics[width=1.\hsize]{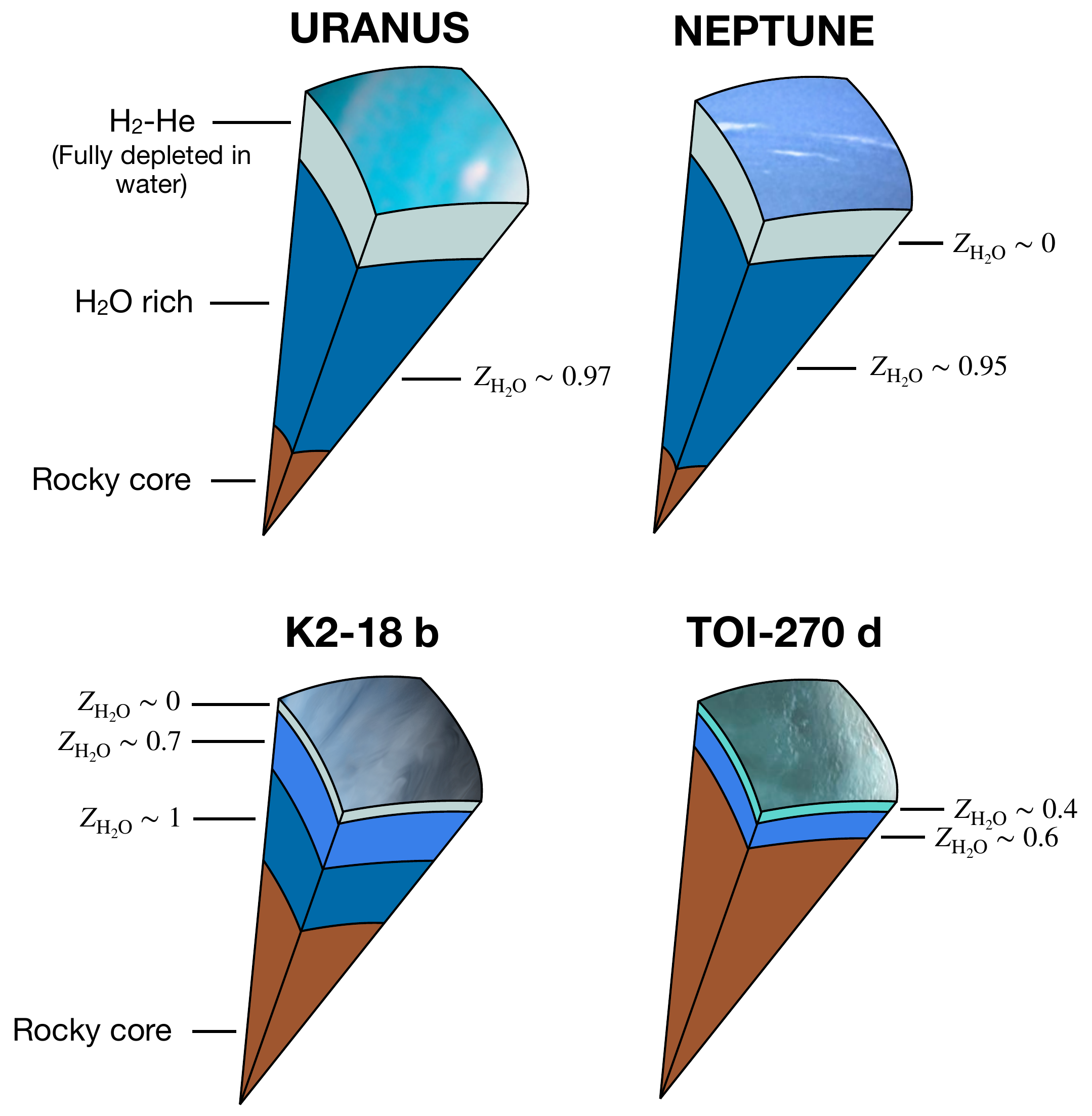}
      \caption{Potential interior structures of Uranus, Neptune, K2-18\,b and TOI-270\,d if they have undergone hydrogen-water demixing. We note that alternatives models are also possible (see Secs.~\ref{sec:solarsystem} and~\ref{sec:exoplanets}). For Uranus and Neptune, the sketches correspond to the cases with $T_{\rm offset}=1100~$K, in which both planets are fully demixed. However, we can't exclude scenarios in which demixing is still ongoing or does not occur at all. The sketch for K2-18\,b shows a mini-Neptune scenario with ongoing demixing, while the one for TOI-270\,d corresponds to a case of full demixing.
      }
         \label{figure:sketchs}
\end{figure}

\begin{acknowledgements}
\texttt{\detokenize{200020_215634}} and the National Centre for Competence in Research ‘PlanetS’ supported by SNSF. We thank S. Glenzer for fruitful discussions.
\end{acknowledgements}
\vspace{-5ex}
%
   \bibliographystyle{aa} 
   \bibliography{biblio} 
%

\begin{appendix} 

\begin{table*}[htbp]
 \centering
 \begin{tabularx}{\textwidth}{lllllllll} \hline\hline
  a & b & c & d & e & f & g & h & i \\\hline
  $1.2035\cdot10^{-4}$ & $0.5501$ & $1.9163\cdot10^{-2}$ & $0.4498$ & $-6.2253\cdot10^{-2}$ & $74.5041$ & $-3.1495\cdot10^{-4}$ & $5.0828\cdot10^{6}$ & $4.0719$ \\\hline\hline
 \end{tabularx}
 \caption{Parameters for Eq. A1.}\label{tab:par}
\end{table*}

\section{Fit}
\label{app:fit}
We found that an equation based on a Lorentzian fits our data and the experimental data the best:
\begin{align}
T_{\rm demix}(P, x_{\rm H_2O}) &=a \times \left[\frac{1}{\pi}\frac{0.5(b+c \times P)}{(x-d)^2+(0.5 \times b)^2}\right] \times \notag \\ 
& (e \times P^3 + f \times P^2 + g \times P + h) + i \times P,
\end{align}
where $x_{\rm H_2O}$ is the number fraction of water. The parameters are given in Tab~\ref{tab:par}.

\section{Water EOS comparison}
\label{app:waterEOS}
The EOS of water remains one of the key sources of uncertainty in modeling the internal structure and evolution of water-rich planets. Despite significant progress through ab initio simulations, most existing EOS models for water are based on density functional theory molecular dynamics (DFT-MD) using the Perdew–Burke–Ernzerhof (PBE) exchange-correlation functional. However, it is well established that PBE tends to systematically underestimate binding energies and overestimate volumes, leading to inaccuracies in the calculated pressure and internal energy, particularly at high pressures and temperatures.

Moreover, current EOS tables often exhibit inconsistencies across different thermodynamic regimes. In particular, the interpolation between low-temperature data (e.g., from experimental measurements or empirical models) and high-temperature DFT-MD results is often poorly constrained. This is especially problematic when bridging the molecular-to-atomic transition or approaching the ideal gas regime, where DFT-MD becomes computationally expensive and alternative models are required. As a result, the thermodynamic consistency—ensuring continuous derivatives and matching of thermodynamic identities—is frequently violated. In addition, the influence of nuclear quantum effects (NQEs), particularly at low temperatures, is often neglected despite their potential impact on phase boundaries and heat capacities.

These uncertainties further justify the introduced shift in temperature $T_{\rm offset}$. In an earlier work of \citet{cano2024}, such a shift was not necessary to produce demixing due to the use of a different water EOS (AQUA) that led to significantly colder adiabats. To compare our adiabats obtained with QEOS to those with AQUA, we calculated two Uranus adiabats with a constant heavy-element mass fraction throughout the envelope of either $Z_{\rm H_2O}=0.4$ or 0.8. These models assumed a core of $4~M_{\oplus}$. Figure~\ref{figure:waterEOS} shows that temperature differences of up to $\sim 1000~$K at a few hundred kbar exist depending on the chosen water EOS. At pressures around a few kbar, the temperature differences are about 100-400~K.

\begin{figure}[h]
   \centering
   \includegraphics[width=0.8\hsize]{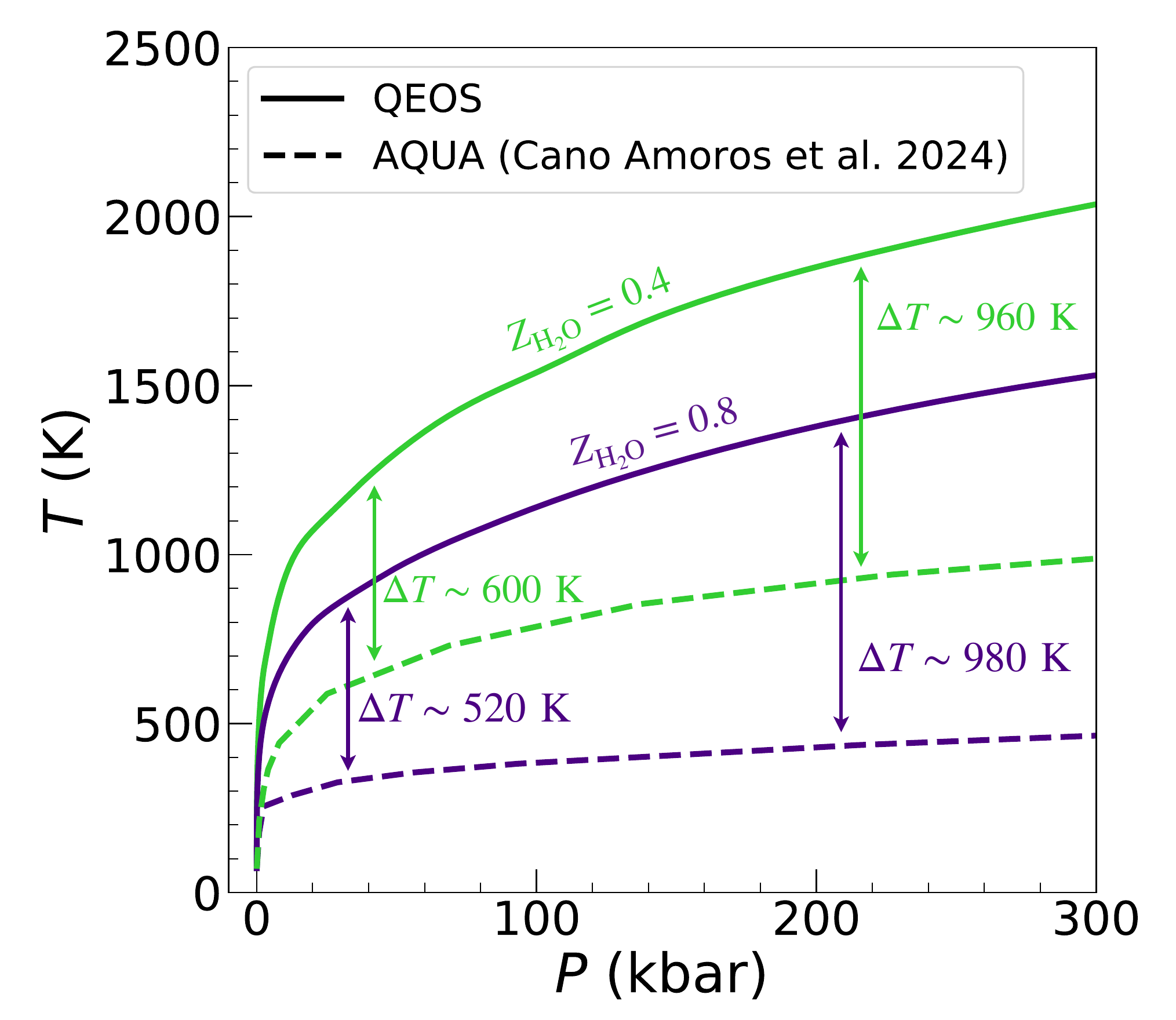}
      \caption{Comparison of Uranus adiabats calculated with different water EOSs. The dashed lines show adiabats from \citet{cano2024}, using AQUA. We calculated adiabats at similar conditions ($T_{\rm 1 bar}=72~$K) using QEOS, shown with solid lines. Colors refer to different assumed water mass fractions in the planetary envelope.
      }
         \label{figure:waterEOS}
\end{figure}

\section{Gravitational moments}
\label{app:J2J4}

Our Uranus models presented in Sec.~\ref{subsec:uranus} showed that hydrogen-water demixing can redistribute material through the planet evolution. This redistribution of material can change the internal density distribution and consequently affect the calculated gravitational moments. We calculated the gravitational moments ($J_2$, $J_4$) of our Uranus models using the Theory of Figures of order 7 implemented by \citet{morf2024}. The results are shown in Fig.~\ref{figure:J2J4}. In the case with $T_{\rm offset}=600~$K, where the outer 2\% by mass are totally depleted in water, $J_2$ changes by about 200~ppm and $J_4$ by about 5~ppm. In the case with $T_{\rm offset}=1100~$K, where the outer 16\% by mass are totally depleted in water, $J_2$ changes by about 1000~ppm and $J_4$ by about 10~ppm. This shows the importance of considering hydrogen-water immiscibility in interior models that try to match the gravity data.

\begin{figure}[h]
   \centering
   \includegraphics[width=\hsize]{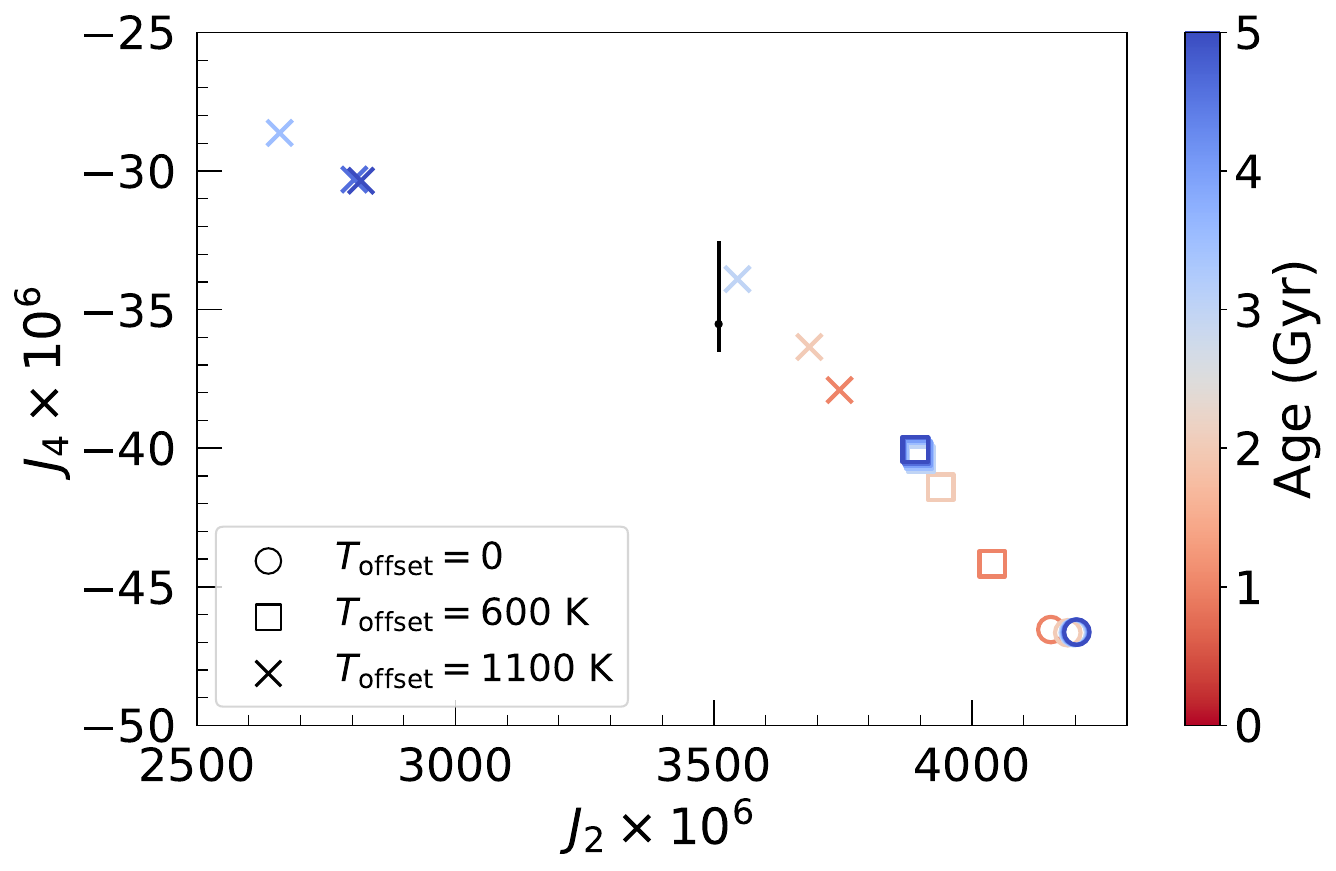}
      \caption{Gravitational moments of our Uranus models. The models correspond to the ones shown in Fig.~\ref{figure:Uranus}. The black errorbar shows the observed values of $J_2$ and $J_4$ from \citet{french2024} which account for the wind corrections from \citet{kaspi2013}, as done in \citet{cano2024}.
      }
         \label{figure:J2J4}
\end{figure}

\section{Liquid water?}
\label{app:liquid}

Figure~\ref{app:liquid} shows a water phase diagram with the adiabats of Uranus and K2-18\,b. Only the portions of the adiabats where water is present are plotted. For Uranus, we show the model with $T_{\rm offset}=1100~$K (see Sec.~\ref{subsec:uranus}). For K2-18\,b, the model corresponds to the mini-Neptune scenario that considers hydrogen-water demixing (see Sec.~\ref{subsec:K2-18b}). For both planets, our models do not suggest the presence of liquid water. The outer part of the envelope that only consists of hydrogen and helium, due to demixing, is too thick to allow the presence of water at low enough pressure and temperature. Further exploration of the parameter space would be needed to definitively rule out the potential presence of liquid water.

\begin{figure}[h]
   \centering
   \includegraphics[width=0.75\hsize]{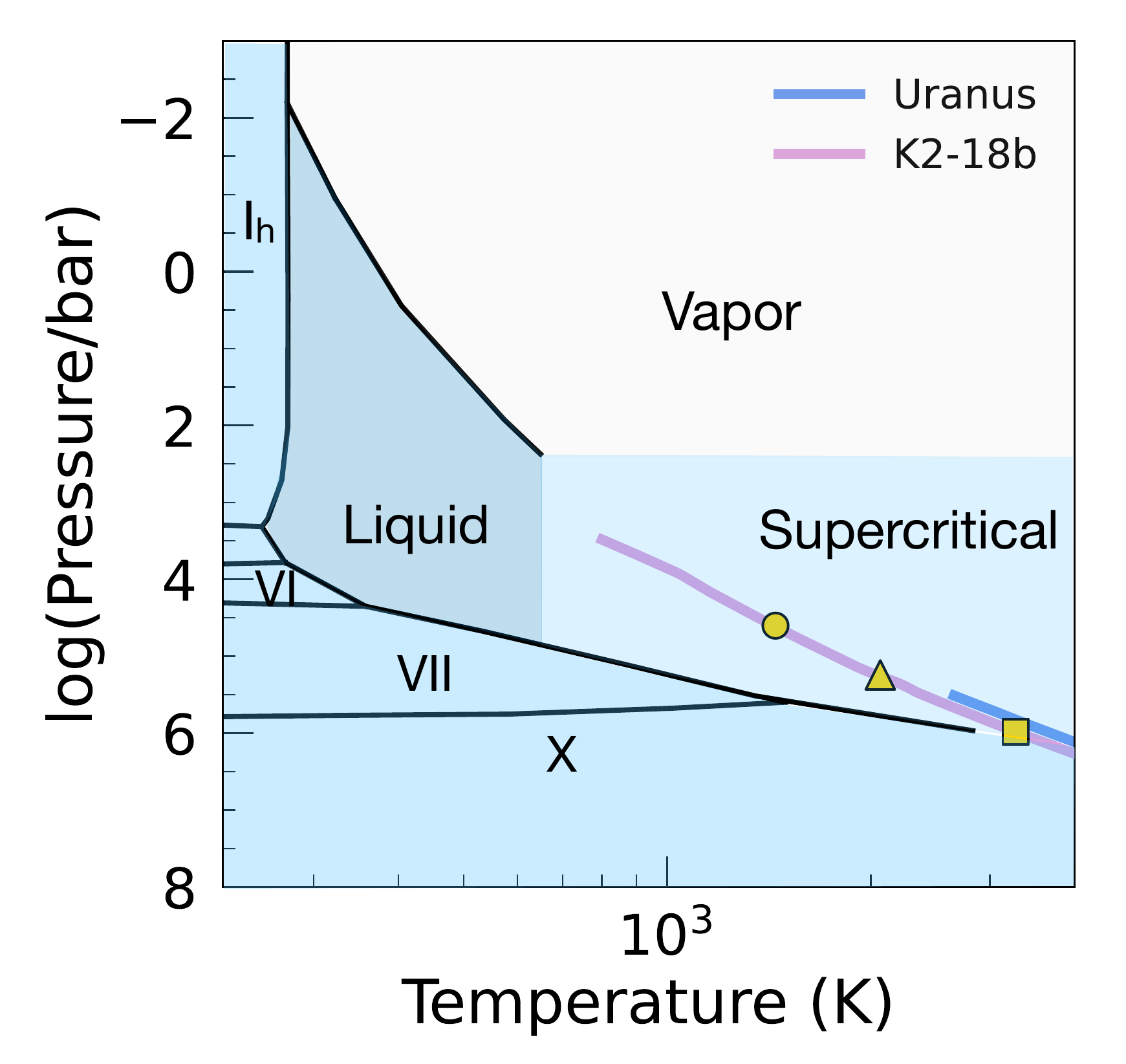}
      \caption{Water phase diagram. The curves for Uranus and K2-18\,b show only the portions of the adiabats where water is present. The yellow circle shows the lower bound of the molecular phase from \citet{redmer2011} while the triangle and square show the expected molecular-to-ionic and ionic-to-superionic transitions in K2-18\,b.
      }
         \label{figure:pasediag_madhu}
\end{figure}

\section{Non-adiabaticity}
\label{app:nonadiabatic}

We calculated a Uranus model with $T_{\rm offset}=600~$K and with $R_{\rho}=0.01$ as defined in Sec.~\ref{sec:discussion}. Figure~\ref{figure:nonadiab} compares this model to the adiabatic case presented in Sec.~\ref{subsec:uranus}. We do not find a significant impact of the superadiabatic gradient on the occurrence of demixing, nor on its effects on the planet’s radius or effective temperature evolution. However, this result depends on the assumed value of $R_{\rho}$ and on the characteristics of the water gradient. In the present model, the water gradient is located at pressures higher than those where phase separation occurs, which may limit its influence on the demixing process.

\begin{figure}[h]
   \centering
   \includegraphics[width=0.85\hsize]{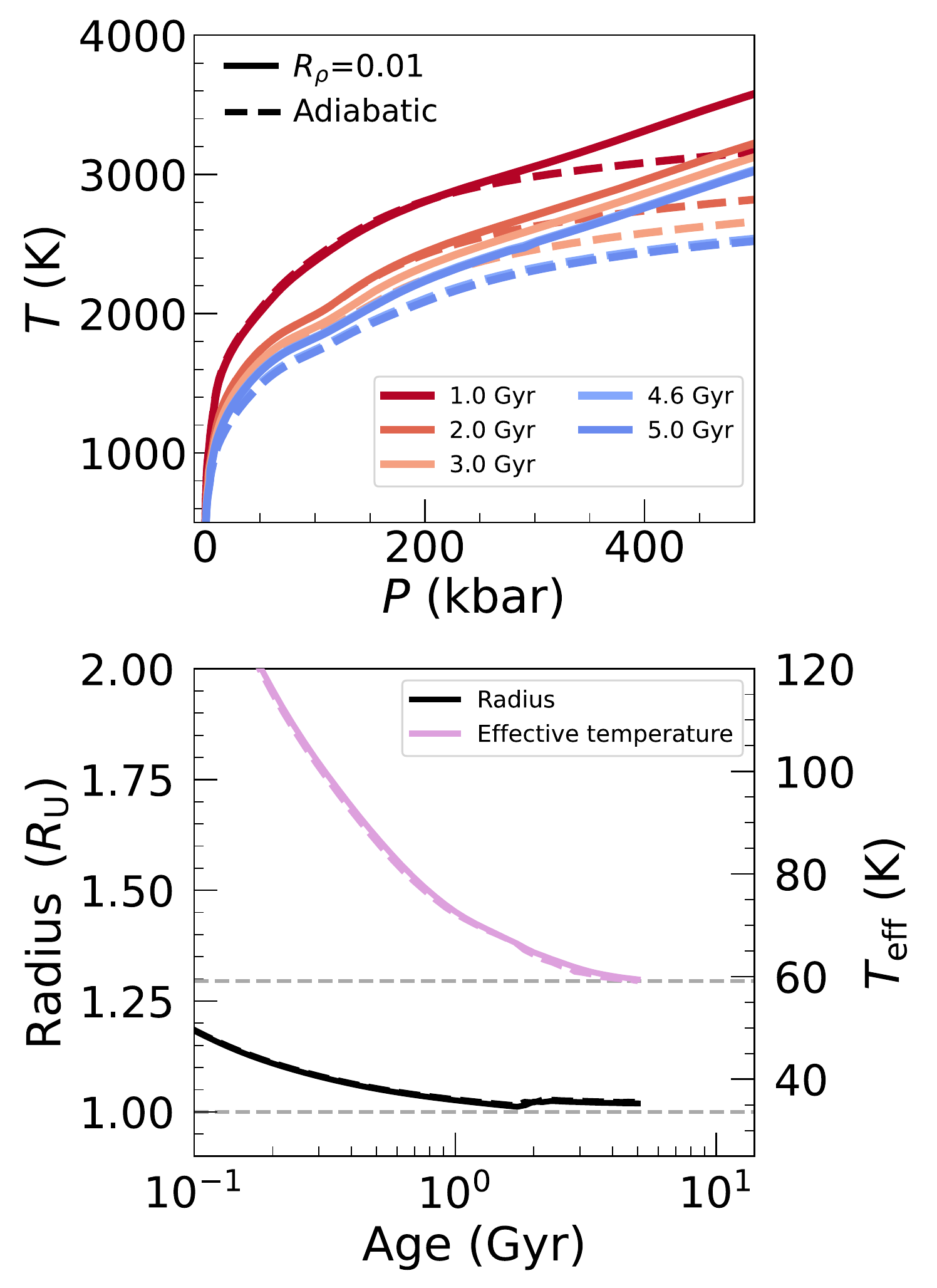}
      \caption{Evolution of Uranus comparing adiabatic and super-adiabatic interiors. \textit{Top panel.} temperature-pressure profiles. \textit{Bottom panel.} radius and effective temperature as a function of age. The horizontal dashed lines show the measured values.
      }
         \label{figure:nonadiab}
\end{figure}

\end{appendix}

\end{document}